\documentstyle[aaspp4,12pt]{article}
\begin{document}

\title{A Semi-Empirical Model of the Infra-Red Universe}
\author{Jonathan C. Tan}
\and
\author{Joe Silk}
\affil{Department of Astronomy\\
University of California\\
Berkeley\\ CA 94720, USA}
\and
\author{Christophe Balland}
\affil{Institut d'Astrophysique Spatiale\\ CNRS, Universite Paris XI\\ Batiment 121, F-91405\\ Orsay, France}

\begin{abstract}
We present a simple model of the infra-red universe, based as much as
possible on local observations. We model the luminosity and number
evolution of disk and starburst galaxies, including the effects of
dust, gas and spectral evolution. Although simple, our approach is
able to reproduce observations of galaxy number counts and the
infra-red and sub-millimeter extra-galactic backgrounds. It provides a
useful probe of galaxy formation and evolution out to high
redshift. The model demonstrates the significant role of the starburst
population and predicts high star formation rates at $z\sim$ 3 to 4,
consistent with recent extinction-corrected observations of Lyman
break galaxies. Starbursting galaxies are predicted to dominate the
current SCUBA surveys. Their star formation is driven predominantly by
strong tidal interactions and mergers of galaxies. This leads to the
creation of spheroidal stellar systems, which may act as the seeds for
disk formation as gas infalls. We predict the present-day baryonic
mass in bulges and halos is comparable to that in disks. From
observations of the extra-galactic background, the model predicts that
the vast majority of star formation in the Universe occurs at
$z\lesssim5$.
\end{abstract}
\newpage
\section{Introduction}

Recent observations of the infra-red (IR) and sub-millimeter (sub-mm)
extra-galactic background (Puget et al 1996; Fixsen et al 1998;
Burigana \& Popa 1998; Schlegel, Finkbeiner \& Davis 1998; Hauser et
al 1998; Dwek et al 1998; Biller et al 1998) provide powerful
constraints on models of galaxy evolution, since a large fraction
($\sim 2/3$) of stellar energy release, after reprocessing by dust, is
emitted in this part of the spectrum. Various number count surveys are
also being pursued across the waveband, from SCUBA at $850\:{\rm \mu
m}$ (Blain et al 1999; Smail et al 1998), to ISO at $175\:{\rm \mu m}$
(Puget et al 1998) and $15\:{\rm \mu m}$ (Aussel et al 1998; Altieri
at al 1998). All of the evidence points to strong evolution in the
galaxy population as we look out to high redshift. We would like to
try and understand this behavior.

Several groups have attempted to model the IR and sub-mm emission from
galaxies, using forward evolution models (Blain \& Longair 1993;
Franceschini et al 1994; Guiderdoni et al 1998; Jimenez \& Kashlinsky
1998; Trentham, Blain and Goldader 1999). Guiderdoni's model is one of
the most detailed. They model structure formation in the Cold Dark
Matter (CDM) cosmological scenario, and introduce star formation with
various recipes for ``quiet-disk'', ``burst'' and ``ultra-luminous''
sources. The relative fractions of these populations and the
timescales of star formation are free parameters in their model. By
tuning these they are able to reproduce current observations, but
there is a lack of physical motivation in many of their assumptions
concerning the number and luminosity evolution of their populations.

In contrast to these models, Malkan \& Stecker (1998) present a
simple, empirically based calculation of the IR background. Using
observed luminosity functions and spectral energy distributions, they
predict reasonable values for the background flux by implementing
various schemes of galaxy evolution. However, these prescriptions are
still without a sound physical basis.

Pei, Fall \& Hauser (1998) present yet a different approach. They
consider the evolution of the global stellar, gaseous, chemical and
radiation contents of the Universe. As inputs they use observations of
the extra-galactic background, interstellar gas density history from
damped Ly$\alpha$ surveys and the rest frame ultra-violet emissivity
history from optical galaxy surveys. They derive cosmic histories of
star formation, metallicity and radiation from stars and dust. Their
global approach cannot distinguish between the different types of star
formation processes in galaxies, such as in disks and starbursts, and
thus comparison to galaxy number count surveys is difficult.

This work approaches the problem in a semi-empirical manner, somewhere
in between the above extremes. The model makes use of reliable
empirical results where they exist. For example, the spectra and
luminosity functions of galaxies in the IR have been well determined
by the IRAS satellite (Saunders et al 1990; Malkan \& Stecker
1998). The star formation rate (SFR) history of the Milky Way's disk
is empirically derived from local observations, and then applied
globally to all disks, assuming our galaxy is typical of these
systems. We present a simple physical model relating SFR history to
the IR luminosity, including dust, gas and spectral evolution. Apart
from disks, observations of IR sources in the local universe, also
reveal many high-luminosity, interacting systems, which appear to be
undergoing an intense burst of star formation (Sanders \& Mirabel
1996). These systems have been termed starbursts. We develop a
physical model for the evolution of this population, assuming they
result from mergers and strong tidal interactions between gas rich
systems. The final theoretical input to the model is the evolution of
the number density and average mass of the disk systems. This is
obtained from the model of collision-induced galaxy formation
(Balland, Silk \& Schaeffer 1998, hereafter BSS98), which attributes
galactic morphology to the number of collisions and tidal interactions
suffered by a particular galaxy.

We evolve the present day populations of disks and starbursts
backwards in time, out to high redshift. This
approach results in relatively few free parameters.  Without fine
tuning, we predict number counts in various wavebands and the IR and
sub-mm extra-galactic background, which agree remarkably well with
observations.  We present the details of the model in \S 2, our
results in \S 3 and our conclusions in \S 4.

\section{The Model}

We model the evolution of two distinct populations which contribute
significant flux in the IR and sub-mm: disk galaxies and
starbursts. The latter we define to be systems undergoing a
violent merger or strong tidal interaction, which leads to high rates
of heavily obscured star formation and the creation of a spheroidal
stellar system or component. An Einstein-de Sitter cosmology
has been assumed with $H_{0}=\:50\:{\rm km\:s^{-1}\:Mpc^{-1}}$. Thus
$t_{0}=13.0$ Gyr. A summary of the model parameters, which are
described below, is shown in Table \ref{tab:param}.

\subsection{Disk galaxies}

We take the present day IR luminosity function of disk galaxies to be
well described by the IRAS far-IR (FIR) luminosity function (Saunders et al
1990) below a certain threshold luminosity, $L_{cut}$:
\begin{equation}
\label{PhiLspi}
\Phi^{disk}(L)=\cases {0 & $L<L_{min}$\cr
C^{disk}(z)(L/L_{*}^{disk}(z))^{1-\alpha}\exp\left[-\frac{1}{2\sigma^{2}}\log^{2}_{10}\left(1+\frac{L}{L_{*}^{disk}(z)}\right)\right] & $L_{min}<L<L_{cut}$\cr
0 & $L>L_{cut}$, \cr}
\end{equation}
where $C^{disk}_{0}=2.6\times 10^{-2}\:h^{3} \:{\rm Mpc^{-3}}$,
$L_{*}^{disk}=10^{8.77}\:h^{-2}\:L_{\odot}$ and $\alpha = 1.09$. We
take $L_{min}=10^{5}\:h^{-2}\:L_{\odot}$. Observationally, the
luminosity function is not well constrained at these low luminosities,
but the extra-galactic background and, for the flux range of interest,
the number counts are insensitive to the choice of $L_{min}$.  We take
the upper limit to be $L_{cut}\sim5.6\times10^{10}\:h^{-2}\:L_{\odot}$
(Sanders \& Mirabel 1996). This defines the difference between
disk-like systems and starbursts, which are modeled in \S 2.2. The
effect of the precise choice of $L_{cut}$ is examined in \S 3. Note,
the above numerical values of $L_{min}$, $L_{*}^{disk}$ and $L_{cut}$
are defined for $z=0$.

Following Malkan \& Stecker (1998), we divide the IRAS luminosity
function into seven different spectral classes, ranging from $10^{-3}$
to $10^{3}\:L_{*}$. The spectra of these classes are derived
empirically from their IRAS colors. These spectra are then
analytically extended beyond 400 ${\rm \mu m}$ into the sub-mm regime,
assuming a dust emissivity dependence of $\nu^{\beta}$, with
$\beta=1.5$ (Franceschini, Andreani, \& Danese 1998; Roche \& Chandler
1993). Results for $\lambda\lesssim1000\:{\rm \mu m}$ are quite
insensitive to the precise choice of $\beta$, since most of the flux
detected in this part of the spectrum is emitted from sources at
$z\gtrsim1$, from their rest frame infra-red.  At shorter wavelengths,
the spectra average over the line and bump features seen around
$10\:{\rm \mu m}$. This may be important for some of the details of
the $15\:{\rm \mu m}$ ISO CAM source counts.  As the luminosities of
the galaxies evolve, the relative proportion of sources in each
spectral class changes, and in this way we take account of spectral
evolution.

Disk galaxies are observed to be undergoing relatively steady star
formation over many Gyr. Their IR luminosity is due to dust heating
from both young stars, which spend most of their lives in dusty
star-forming regions, and older stars, which contribute to the general
interstellar radiation field (ISRF). The present-day fraction of the
total from young star heating is $f_{0}^{young}\sim0.65$ (Mayya \& Rengarajan
1997; Devereux et al 1994; Xu \& Helou 1996; Walterbos \& Greenawalt
1996). We consider the luminosity evolution due to these two heating
sources separately.

The IR luminosity of disks due to young stars, $L^{ir,young}$, is
expected to evolve with redshift in the following manner:
\begin{equation}
\label{Lirsfr}
\frac{L^{ir,young}(z)}{L^{ir,young}_{0}}=\frac{\phi(z)}{\phi_{0}}\frac{(1-e^{-\tau^{uv}(z)})}{(1-e^{-\tau^{uv}_{0}})}\frac{M^{gal}(z)}{M^{gal}_{0}},
\end{equation}
where $\phi(z)$ is the local star formation rate (SFR) per unit disk
mass of disk galaxies, $\tau^{uv}(z)$ is the disk optical depth to
young stellar energy release and $M^{gal}(z)$ is the mean baryonic
mass of disk galaxies. These are all redshift dependent.
 
\begin{figure}
\plotone{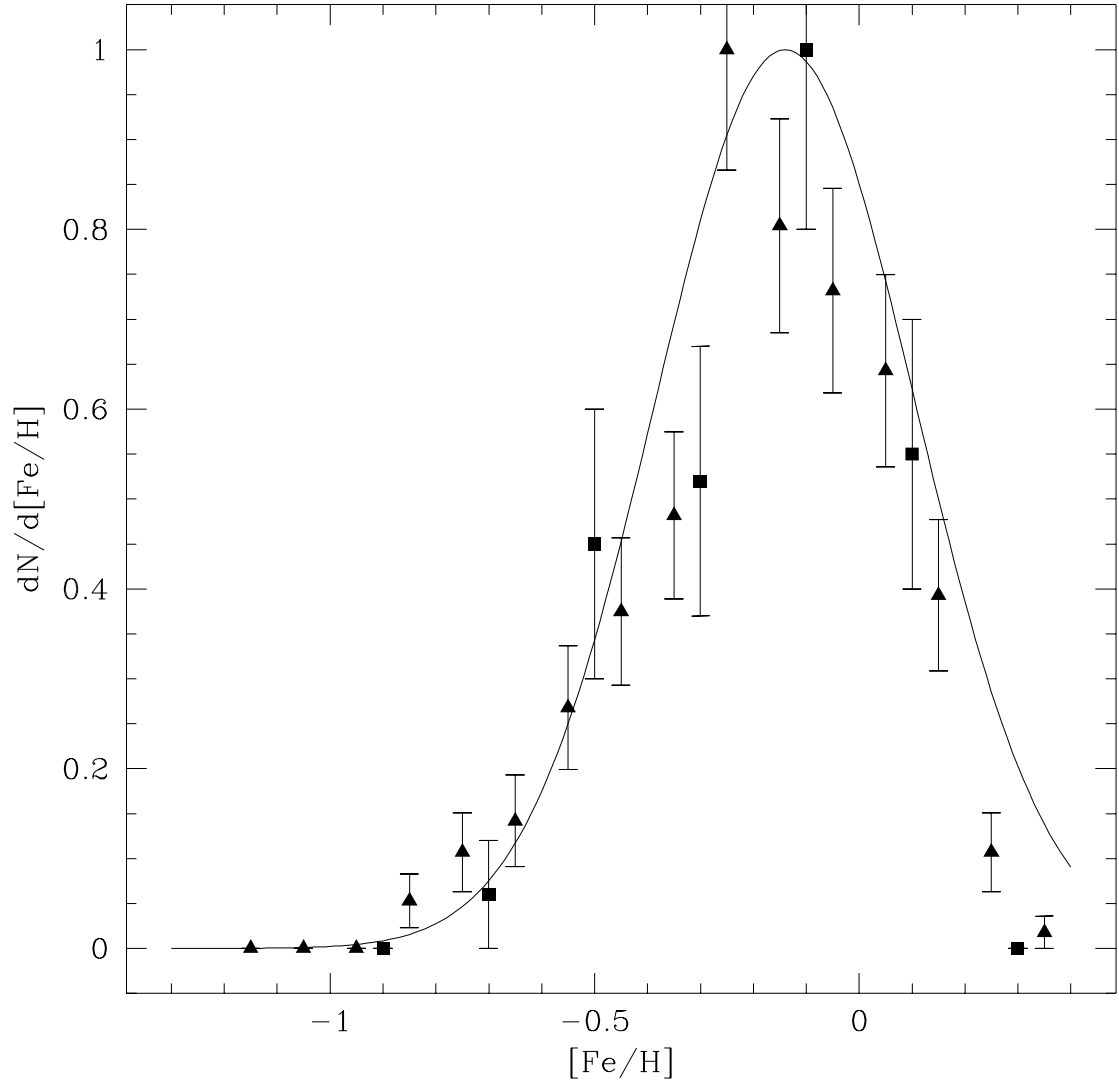}
\caption{
Metallicity distributions of G-type stars in the solar neighborhood. Data from Rocha-Pinto \& Machiel (1996, {\it triangles}) and Wyse \& Gilmore (1995, {\it squares}). {\it Solid curve:} Gaussian fit to data - the SFR history model depends only on the fit for [Fe/H]$<0$.
\label{fig:nummetals}}
\end{figure}

\begin{figure}
\plotone{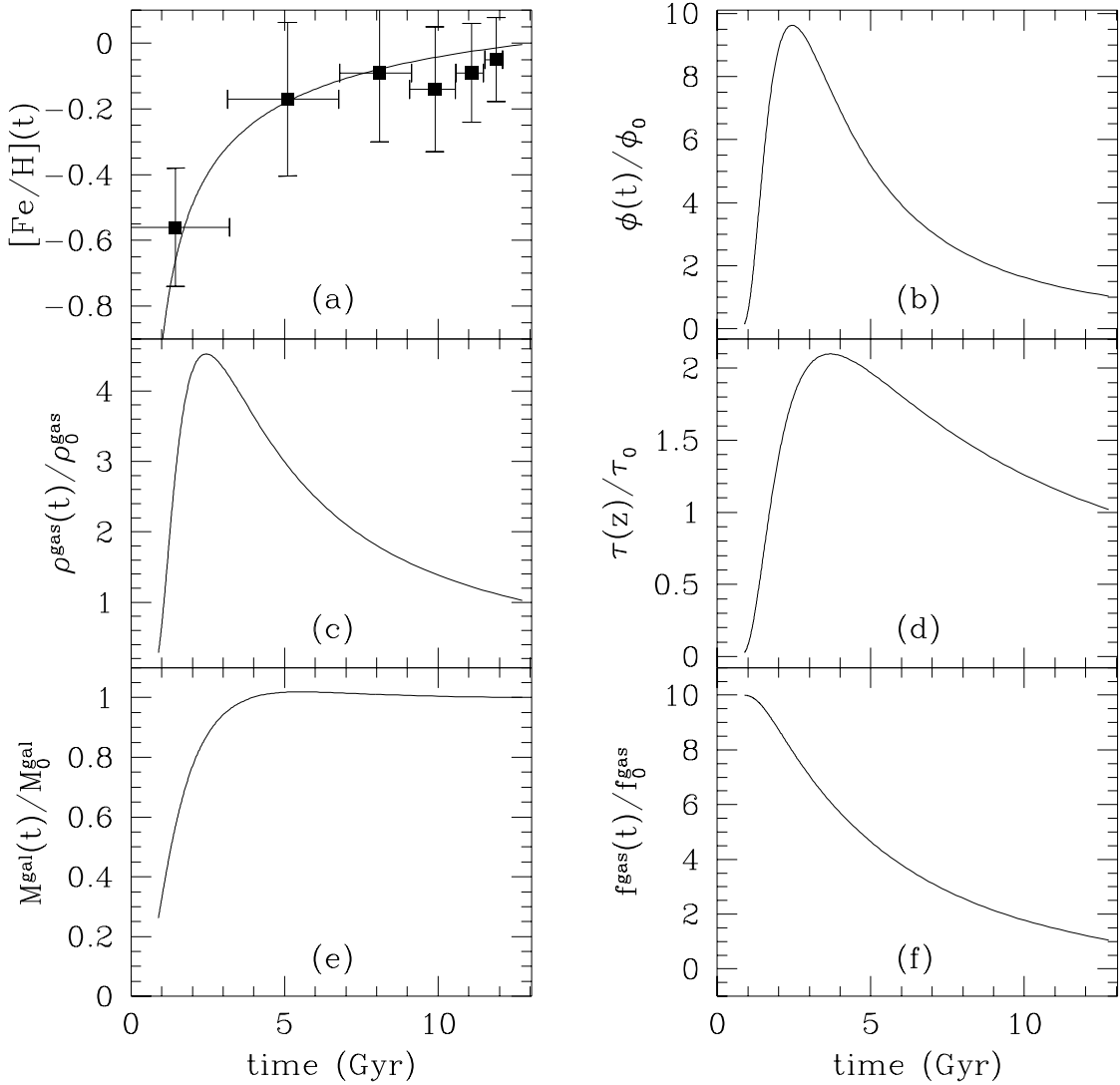}
\caption{
Disk Galaxies: Disk star formation is assumed to start 12 Gyr ago, with $t_{0}=13\:{\rm Gyr}$. This corresponds to $z\approx 5.5$ for our assumed cosmology.
(a) Metallicity History of solar neighborhood. Data from Edvardsson et al (1993). {\it Solid curve:} Fit to data, with the additional constraint of the present-day stellar disk surface density.
(b) Relative SFR.
(c) Relative gas density in the disk star forming region, assuming a Schmidt law of index $n=1.5$.
(d) Relative optical depth.
(e) Relative mean galactic mass (from BSS98).
(f) Relative gas fraction of the total baryonic matter bound to the galaxy.
\label{fig:diskmodel}}
\end{figure}

We derive $\phi$ from the metallicity distributions, $dN/dZ$, of local
G-type stars (Rocha-Pinto \& Maciel 1996; Wyse \& Gilmore 1995),
(Figure \ref{fig:nummetals}), and the age-metallicity relationship,
$dZ/dt$, of local F-type stars (Edvardsson et al 1993), (Figure
\ref{fig:diskmodel}a), using
\begin{equation}
\label{philocal}
\phi(t)\propto\frac{dN}{dZ}\frac{dZ}{dt}.
\end{equation}
$\phi(t)$ is further constrained by having to produce a local
present-day stellar disk surface density of $\sim 40\:{\rm
M_{\odot}\:pc^{-2}}$ (Sackett 1997), assuming the returned mass
fraction is small.  From a sample of nearby white dwarfs and
consideration of their cooling curves, Oswald et al (1996) find the
local disk of the Milky Way has an age of $\sim10\pm1$ Gyr. Using a
similar method, Knox, Hawkins \& Hambly (1999) derive an age of
$10_{-1}^{+3}$ Gyr. However, because we wish to describe the galaxy
averaged evolution of disks, we take the mean age to be $\sim12$
Gyr, since it is expected the inner disk will start forming at earlier
times. For simplicity we use this formation time to complete the
normalization of $\phi(t)$, which we now take to approximate the
evolution of the total Galactic SFR history. An age of 12 Gyr
corresponds to a disk formation time, $t_{f}\approx1$ Gyr, and
redshift, $z_{f}\approx5.5$. At these high redshifts, the model
predictions of the source counts and backgrounds are quite insensitive
to the precise choice of $z_{f}$. The major uncertainty in this
procedure is in fitting the age-metallicity relation, because of the
large scatter in the data. We fit the data for $dN/dZ$, which show
less scatter, with a Gaussian. Only the fit at solar metallicities and
below affects the derived SFR history. $dZ/dt$ is fit with a function
which rises asymptotically from low metallicities at $t\sim t_{f}$,
and then levels off to solar metallicity at $t\sim t_{0}$, with the
additional constraint that the integrated SFR history (Figure
\ref{fig:diskmodel}b) matches the observed stellar disk surface
density. The SFR rises rapidly at early times, peaks at a level
roughly ten times that of today, and then decays exponentially.
Assuming the Milky Way is typical of disk galaxies, we apply $\phi$ to
the entire population of disks. This method enables us to empirically
probe the evolution of disk systems separately from the other
components of the Universe.

With a redshift independent dust-to-metals mass ratio (Pei et al 1998,
their Figure 1), $\tau$ is proportional to the metallicity, $Z$, and
the local gas density in the disk, $\rho^{gas}$. Thus,
\begin{equation}
\label{tau}
\frac{\tau(z)}{\tau_{0}}=\frac{Z(z)}{Z_{0}}\frac{\rho^{gas}(z)}{\rho_{0}^{gas}}=\frac{Z(z)}{Z_{0}}\left(\frac{\phi(z)}{\phi_{0}}\right)^{2/3},
\end{equation}
where we have used the Schmidt law, $\phi\propto(\rho^{gas})^{n}$,
with $n=1.5$, to relate gas density to the rate of star
formation. This relation is observed to hold over a large range of gas
densities relevant to both disks and starbursts (Kennicutt 1998). The
gas density history is shown in Figure \ref{fig:diskmodel}c. We know
the metallicity history directly from the age-metallicity relationship
(Figure \ref{fig:diskmodel}a). The optical depth history is shown in
Figure \ref{fig:diskmodel}d.

$M^{gal}(z)$ is derived from the model of collision-induced galaxy
formation of BSS98 and is shown in Figure \ref{fig:diskmodel}e. The
mean baryonic mass in a galaxy is controlled by the competition
between merging and cooling.  It results from consideration of the
Press-Schechter mass function, together with an applied cooling
constraint, which gives an upper limit to galaxy masses. There is
little evolution in the mean mass out to $z\sim2$. By $z\sim 5$ it has
fallen by a factor of a few.

The fraction of the baryonic mass bound to a galaxy that is in the
form of gas, $f^{gas}$, can be calculated, given $\phi$ and
$f^{gas}_{0}$.  This gas is not necessarily in the disk of the galaxy,
but may be in the process of falling in from the halo.  It is
available to form stars in the event of a merger or close encounter
with another galaxy, as tidal forces will channel the gas (or some
portion of it) into the central starbursting regions. Thus knowledge
of the evolution of $f^{gas}$ is necessary for modeling of the
starbursts in \S 2.2. From observations of local disk galaxies, we
take $f^{gas}_{0}\sim 0.1$ (Young \& Scoville 1991; Young et al
1995). The effect of varying this choice is investigated in \S3. Its
evolution is shown in Figure \ref{fig:diskmodel}f.

We describe the IR luminosity due to dust heating by the older stellar
population as follows:
\begin{equation}
\label{Lirold}
\frac{L^{ir,old}(z)}{L^{ir,old}_{0}}=\frac{\int_{t_{f}}^{t_{z}}\phi(t)\:dt}{\int_{t_{f}}^{t_{0}}\phi(t)\:dt}\frac{(1-e^{-\tau^{opt}(z)})}{(1-e^{-\tau_{0}^{opt}})}\frac{M^{disk}(z)}{M^{disk}_{0}}.
\end{equation}
For simplicity it has been assumed that the intensity of the ISRF of
disks in the optical scales as the integrated SFR history. The
combination of $L^{ir,young}$ and $L^{ir,old}$ is a simple
representation of the true situation in which there is a continuously
varying contribution to dust heating over the whole mass spectrum and
environments of stars present in a galaxy.

From Wolfire et al (1999) we set $\tau^{uv}_{0}\sim 0.7$. We assume
$\tau^{opt}_{0}\sim 0.1$, placing present-day disks in the optically
thin limit, with respect to the general ISRF. The calculation of the
IR and sub-mm number counts and background is relatively insensitive
to the choices of $\tau^{uv}_{0}$ and $\tau^{opt}_{0}$, provided we
are in the optically semi-thick and thin regimes respectively, since,
in this model, it is the scaling of the luminosity function, with
respect to today's, which is important.

\begin{figure}
\plotone{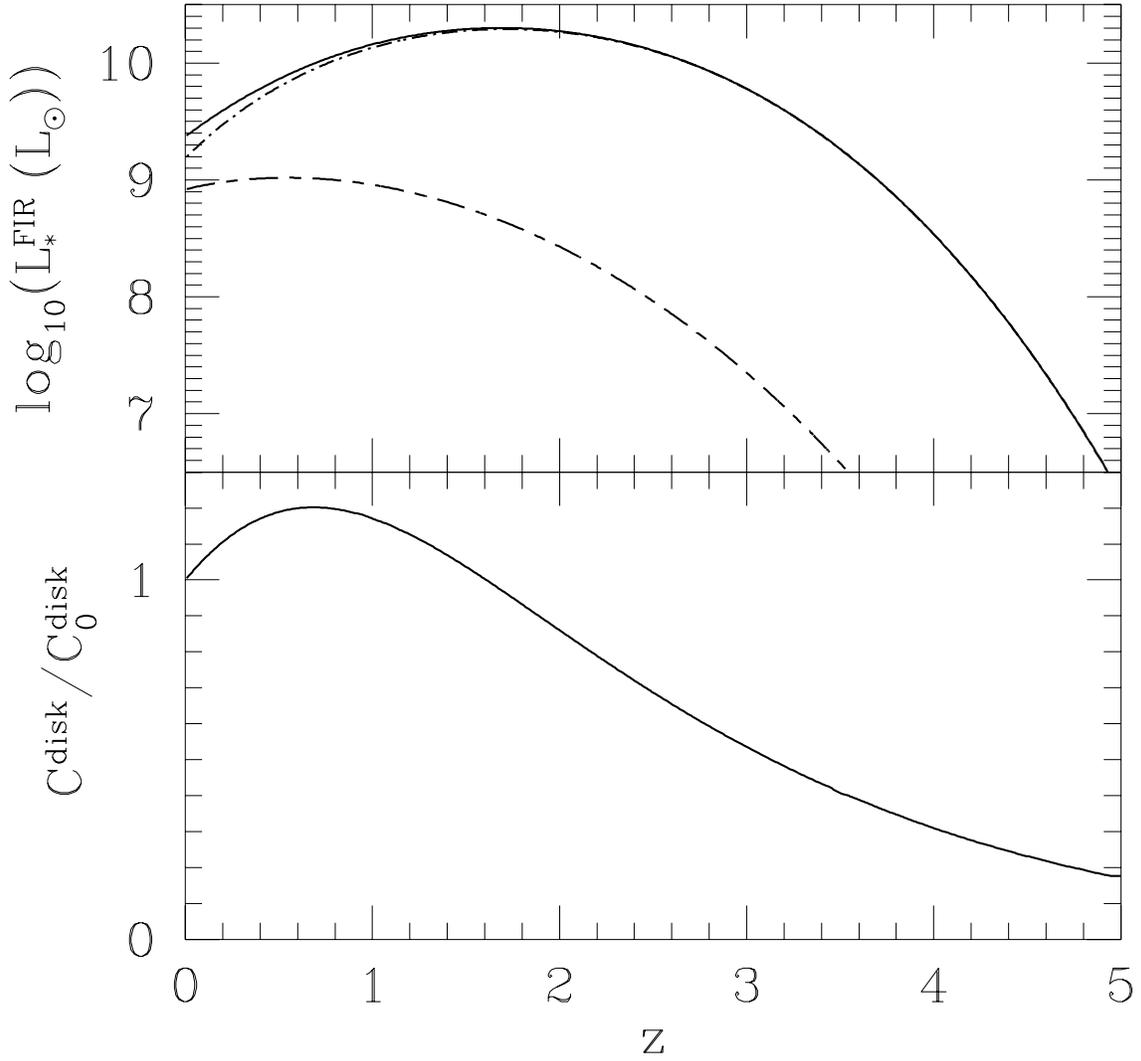}
\caption{
Evolution of the Disk Galaxy far-IR Luminosity Function: (a) ({\it Top Panel}) Luminosity evolution for total $L_{*}^{disk}$ ({\it solid line}), $L^{young}_{*}$ ({\it dot - dash line}) and $L^{old}_{*}$ ({\it short dash - long dash}).
(b) ({\it Bottom Panel}) Number density evolution for $C^{disk}$ from BSS98.
\label{fig:diskevol}}
\end{figure}

We implement luminosity evolution of the present-day characteristic
luminosity, $L^{disk}_{*}$, using equations (\ref{Lirsfr}) and
(\ref{Lirold}), each contributing the appropriate fraction,
$f_{0}^{young}$ and $1-f_{0}^{young}$ respectively to the total. This
is shown in Figure \ref{fig:diskevol}a.

We obtain number density evolution (Figure \ref{fig:diskevol}b) from
the models of BSS98 for field (i.e. non-cluster) galaxies, and apply
this scaling to the normalization constant, $C^{disk}(z)$, of the
luminosity function. In their model disk galaxies form from clouds
which experience few, if any, collisions between the formation time
and the epoch under consideration. Strong collisions will tend to
prevent the gas from settling into a disk, allow for tidal exchanges
which average out angular momentum, and lead to the formation of
ellipticals.

\subsection{Starbursts}

We model starburst galaxies as gas rich systems which are undergoing a
merger or a strong tidal interaction. We assume the starburst number
density scales with the rate of such interactions, while the mean
luminosity scales as the average galactic mass, $M^{gal}$, and the gas
fraction of baryonic matter, $f^{gas}$, available for star
formation. We treat the star formation as being completely obscured by
dust. Observations support this approximation (Sanders \& Mirabel
1996), and it is imagined that even for metal free systems, the
initial burst-induced star formation acts very quickly to pollute the
ISM and so obscure the vast majority of the stellar energy release.

We model the local ($z\lesssim0.2$) starburst population with the high
luminosity end of the IRAS FIR luminosity function:
\begin{equation}
\label{PhiSB}
\Phi^{sb}(L)=\cases {0 & $L<L_{cut}$\cr
C^{sb}(z)(L/L_{*,0}^{sb}(z))^{1-\alpha}\exp\left[-\frac{1}{2\sigma^{2}}\log^{2}_{10}\left(1+\frac{L}{L_{*,0}^{sb}(z)}\right)\right] & $L_{cut}<L<L_{max}$\cr
0 & $L>L_{max}$,\cr}
\end{equation}
where $C^{sb}_{0}=2.6\times 10^{-2}\:h^{3} \:{\rm Mpc^{-3}}$,
$L_{*}^{sb}=10^{8.77}\:h^{-2}\:L_{\odot}$, $\alpha = 1.09$ (Saunders
et al 1990) and $L_{cut}\sim5.6\times10^{10}\:h^{-2}\:L_{\odot}$
(Sanders \& Mirabel 1996). We take
$L_{max}=10^{15}\:h^{-2}\:L_{\odot}$. The results are insensitive to
this choice because of the steepness of the luminosity function,
equivalent to $L^{-2.35}$, (Kim \& Saunders 1998). Note, $L_{*}^{sb}$
is not a characteristic starburst luminosity. It is simply used to
parameterize the luminosity function.

We use the same spectral luminosity classes as for the disk
model. Starbursts, with higher FIR luminosities than disks, have a
stronger component of warm dust emission.

$L_{*}^{sb}$ is modeled to evolve as follows:
\begin{equation}
\label{LstarSB}
\frac{L_{*}^{sb}(z)}{L_{*,0}^{sb}}=\frac{M^{gal}(z)}{M_{0}^{gal}}\frac{f^{gas}(z)}{f_{0}^{gas}}.
\end{equation}
Since all the energy released by star formation is reradiated in the
far-IR, $L_{*}^{sb}$ is proportional to the average galactic mass and
gas fraction of the merging systems.

\begin{figure}
\plotone{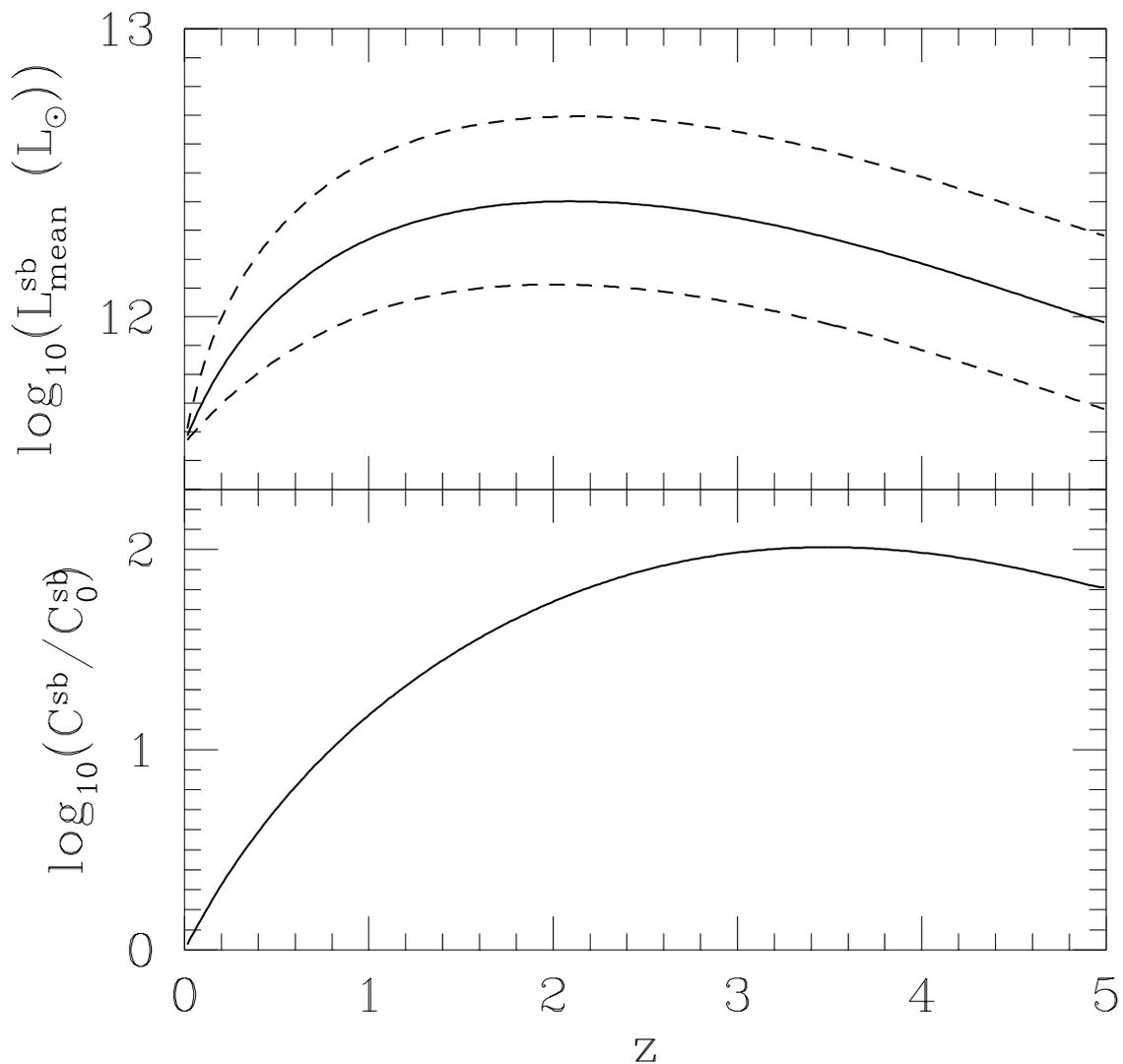}
\caption{ Evolution of the Starburst far-IR Luminosity Function: (a)
({\it Top Panel}) Luminosity evolution of the mean starburst
luminosity, $L_{mean}^{sb}$. Model results: {\it solid line} -
fiducial model $f_{0}^{gas}=0.1$, {\it upper dashed line} -
$f_{0}^{gas}=0.05$, {\it lower dashed line} - $f_{0}^{gas}=0.2$. (b)
({\it Bottom Panel}) Number density evolution for $C^{sb}$.
\label{fig:SBevol}}
\end{figure}

To model the number density evolution, we assume $C^{sb}$ is
proportional to the rate of collisions between gas rich systems,
$\Gamma^{coll}$. The collision rate of any given galaxy is
\begin{equation}
\label{gamma1}
\Gamma^{coll}_{1}=\frac{v}{\lambda_{mfp}}\propto n^{gal}R^{2}v\propto n^{gal}_{c}(1+z)^{3}M^{gal}\frac{v_{0}}{(1+z)^{1/2}},
\end{equation}
where $v$ is the mean peculiar velocity of galaxies, $\lambda_{mfp}$
is the mean free path before a collision or strong interaction occurs,
$R$ is the mean galactic linear size, $n^{gal}$ is the true number density of
(gas rich) galaxies and $n^{gal}_{c}$ is the corresponding comoving
number density. We assume that galaxies are disk-like so that
$M^{gal}=k R^{2}$ for some constant $k$, which we take to be redshift
independent. The redshift dependence of $v=v_{0}(1+z)^{-1/2}$ results
from consideration of structure formation in the linear regime. The
total collision rate is thus:
\begin{equation}
\label{gammaSB}
\frac{\Gamma^{coll}}{\Gamma^{coll}_{0}}=\frac{C^{sb}}{C^{sb}_{0}}=(1+z)^{5/2}\left(\frac{n_{c}^{gal}}{n_{c,0}^{gal}}\right)^{2}\frac{M^{gal}}{M^{gal}_{0}}.
\end{equation}
We obtain $n_{c}^{gal}(z)$ from BSS98, by summing their disk and
irregular morphological types. These dominate the galaxy population at
any particular epoch.  The luminosity and number density evolutions of
the starburst population are shown in Figure \ref{fig:SBevol}. The luminosity evolution depends on the value of $f_{0}^{gas}$ and the effect of its variation is shown.

\section{Results}

For our IR and sub-mm sources we predict number counts, redshift
distributions, the intensity of the extra-galactic background and the
global SFR history. The values of our model's parameters are based
directly on observations (Table \ref{tab:param}) and so we do not
attempt to fine tune them to obtain ``perfect'' fits to all the
observations. Rather, we demonstrate that this simple model of galaxy
evolution is consistent with all the available data, and examine the
effects of varying our two most sensitive model parameters, $L_{cut}$
and $f_{0}^{gas}$.

\begin{figure}
\plotone{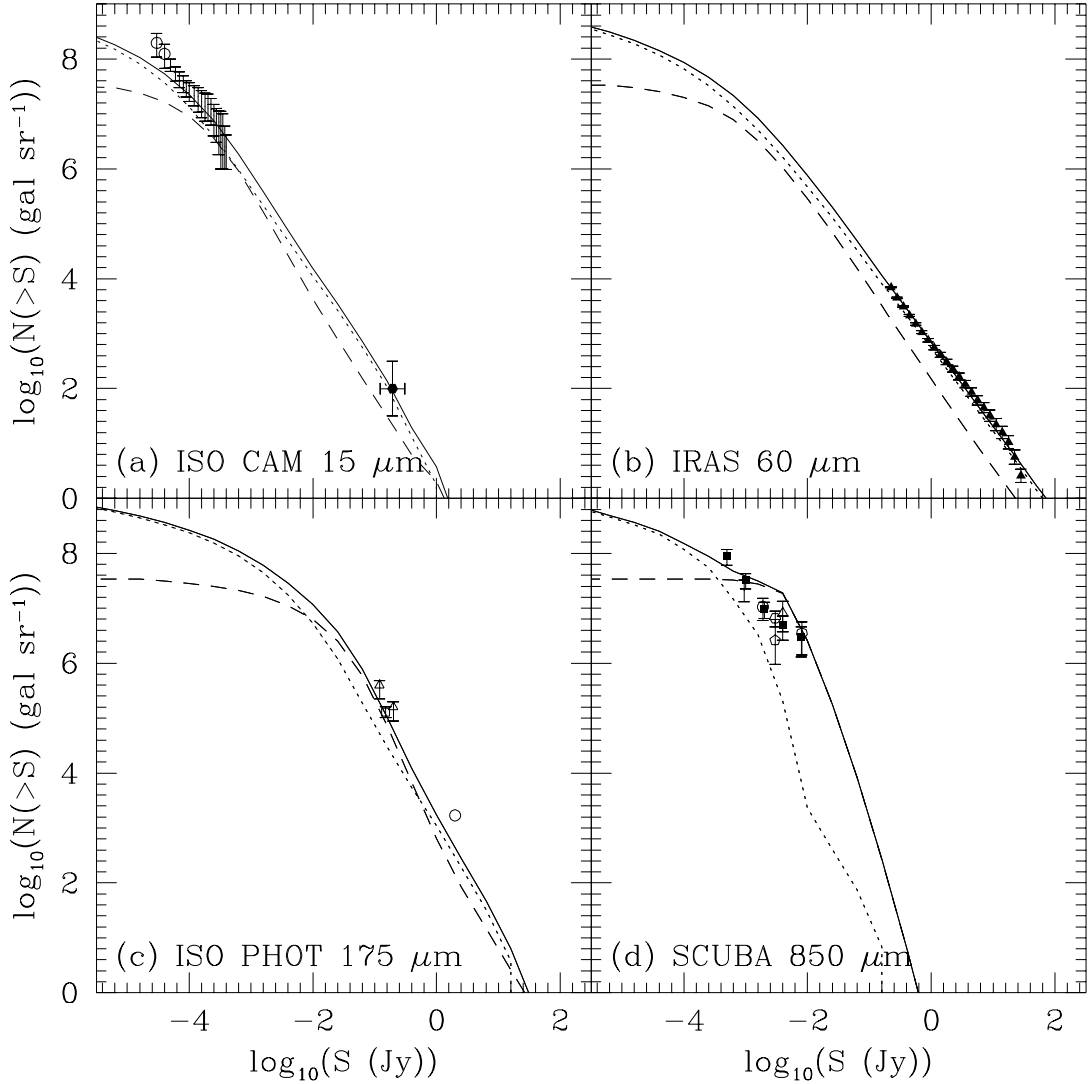}
\caption{
Integrated number counts. Model results: {\it dotted} - disks, {\it dashed} - starbursts, {\it solid} - total.
(a) ISO CAM $15 \:{\rm \mu m}$. Data: {\it solid hexagon} - Rush et al (1993) (IRAS $12 \:{\rm \mu m}$),
{\it vertical lines} - Aussel et al (1998) (ISO CAM) and {\it open circles} - Altieri et al (1998) (ISO CAM). 
(b) IRAS $60 \: {\rm \mu m}$. Data: {\it solid triangles} - Lonsdale et al (1990). 
(c) ISO PHOT $175 \: {\rm \mu m}$. Data: {open triangles} - Puget et al (1998), {open square} - Kawara et al (1998) (raw number counts) and {\it open circle} - Stickel et al (1998) (serendipity survey).
(d) SCUBA $850 \:{\rm \mu m}$. Data: {\it solid circles} - Blain et al (1999) (SCUBA lens survey) and {\it open symbols} - compiled in Smail et al (1998).
\label{fig:numcounts}}
\end{figure}

The disk, starburst and total integrated number counts are shown in
Figure \ref{fig:numcounts}.  The model agrees well with observations
made in the IRAS $12 \:{\rm \mu m}$ (Rush et al 1993), ISO CAM $15
\:{\rm \mu m}$ (Aussel et al 1998; Altieri et al 1998), IRAS $60
\:{\rm \mu m}$ (Lonsdale et al 1990), ISO PHOT $175 \:{\rm \mu m}$
(Puget et al 1998) and SCUBA $850 \:{\rm \mu m}$ (Blain et al 1999;
Smail et al 1998 and references therein) passbands. At 15 ${\rm \mu
m}$ disks dominate the counts at high fluxes, while the starburst
contribution becomes comparable between $10^{-3}$ to $10^{-4}$ Jy. At
lower fluxes, the starburst counts flatten off, as we are probing the
limit of the distribution. At 60 ${\rm \mu m}$ the disks again
dominate at the high flux end, with starbursts becoming more important
at around $10^{-2}$ Jy. At 175 ${\rm \mu m}$ the starburst counts rise
steeply to dominate the total between $\sim 1$ and $10^{-2}$ Jy, which
is the region probed by current ISO PHOT observations. At 850 ${\rm
\mu m}$ starbursts completely dominate at all fluxes above $10^{-3}$
Jy.

\begin{figure}
\plotone{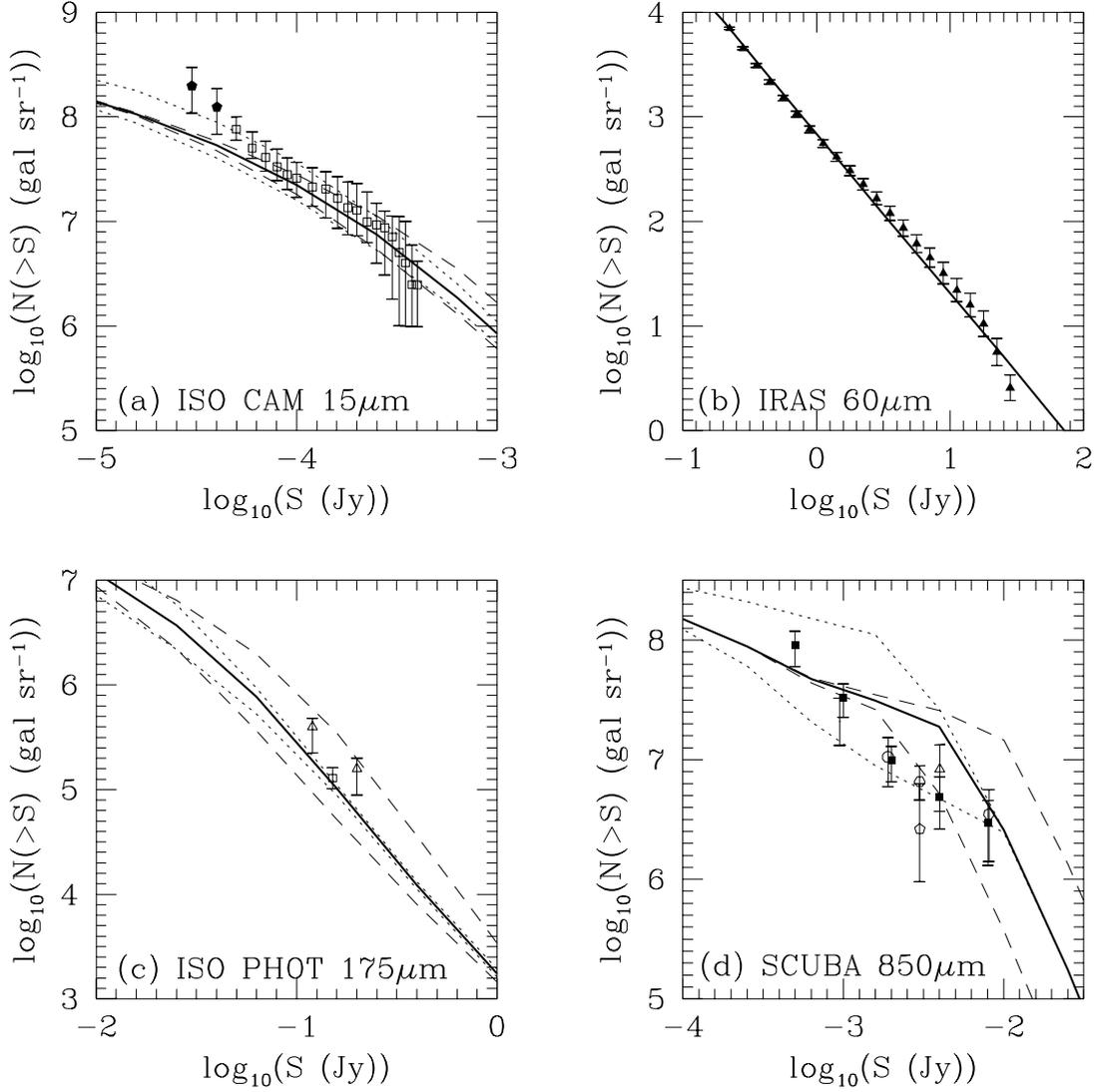}
\caption{
Integrated number counts. Model results: {\it heavy solid line} - fiducial model: $L_{cut}=5.6\times 10^{10}\:h^{-2}\:L_{\odot}$ and $f_{0}^{gas}=0.1$, {\it upper dotted line} - $L_{cut}=11.2\times 10^{10}\:h^{-2}\:L_{\odot}$, {\it lower dotted line} - $L
_{cut}=2.8\times 10^{10}\:h^{-2}\:L_{\odot}$, {\it upper dashed line} - $f_{0}^{gas}=0.05$, {\it lower dashed line} - $f_{0}^{gas}=0.2$.
(a) ISO CAM $15 \:{\rm \mu m}$. Data: {\it open squares} - Aussel et al (1998) (ISO CAM) and {\it open circles} - Altieri et al (1998) (ISO CAM). 
(b) IRAS $60 \: {\rm \mu m}$. Data: {\it solid triangles} - Lonsdale et al (1990). 
(c) ISO PHOT $175 \: {\rm \mu m}$. Data: {open triangles} - Puget et al (1998) and {open square} - Kawara et al (1998) (raw number counts).
(d) SCUBA $850 \:{\rm \mu m}$. Data: {\it solid circles} - Blain et al (1999) (SCUBA lens survey) and {\it open symbols} - compiled in Smail et al (1998).
\label{fig:numcountszoom}}
\end{figure}

Figure \ref{fig:numcountszoom} shows the total counts in expanded
regions of the flux-number count diagrams relevant to the latest
observations. The ISO 15 ${\rm \mu m}$, ISO 175 ${\rm \mu m}$ and
SCUBA 850 ${\rm \mu m}$ plots also show the effect of varying
$f_{0}^{gas}$ and $L_{cut}$ from the fiducial values. Varying
$f_{0}^{gas}$ affects the starburst population while $L_{cut}$
dictates the relative contribution of disks to starbursts. The IRAS 60
${\rm \mu m}$ data, sampling the low redshift population, are very
insensitive to these parameters.

\begin{figure}
\plotone{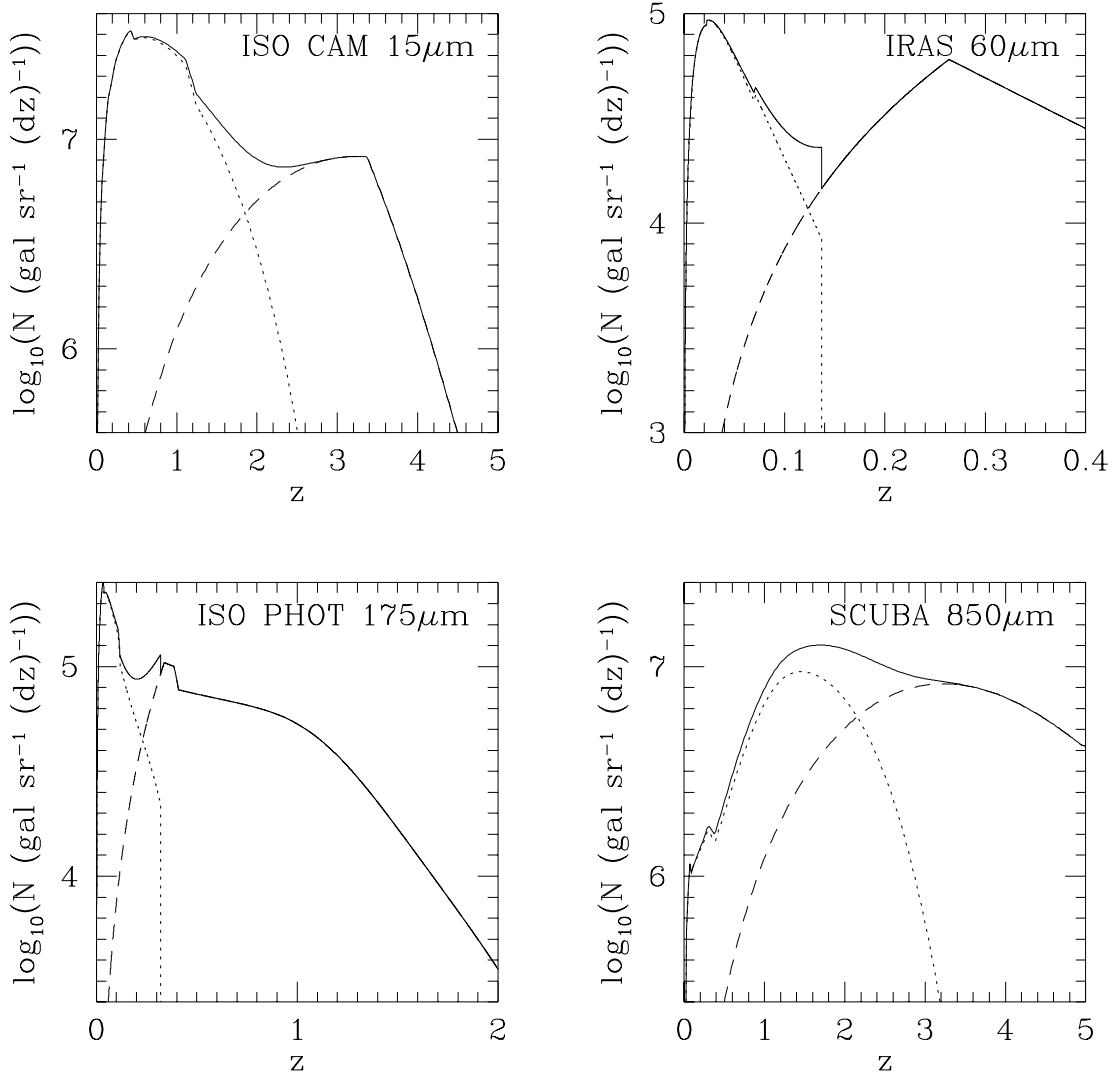}
\caption{
Redshift distributions. Model results: {\it dotted} - disks, {\it dashed} - starbursts, {\it solid} - total. The discontinuities are artifacts caused by the simplistic division of the luminosity function into disk galaxies and starbursts and discrete spectral classes. 
{\it top left:} ISO CAM $15 \:{\rm \mu m}$ for $S>40\:{\rm \mu Jy}$.
{\it top right:} IRAS $60 \:{\rm \mu m}$ for $S>0.20$ Jy.
{\it bottom left:} ISO PHOT $175 \:{\rm \mu m}$ for $S>0.16$ Jy.
{\it bottom right:} SCUBA $850 \:{\rm \mu m}$ for $S>0.63$ mJy.
\label{fig:zdist}}
\end{figure}

Figure \ref{fig:zdist} shows the predicted redshift distributions of
the source populations observed by ISO CAM, IRAS, ISO PHOT and
SCUBA. The discontinuities in the distributions are artifacts
resulting from our simplistic method of dividing the luminosity
function into disk and starburst sources and discrete spectral
classes. The observed disks are at lower redshifts relative to the
bulk of the observed starbursts. In the case of the SCUBA sources
detected with $S>0.63$ mJy, about $2/3$ are predicted to be
starbursting systems, and the rest normal disks. Most of these
sources are predicted to be at redshifts greater than one, with many
seen out to redshift five and beyond. This is due to the steep slope
of the Rayleigh-Jeans portion of the modified blackbody spectra of the
sources. As more redshift determinations are made of these source
samples, it will be interesting to compare the inferred distributions,
corrected for clustering effects and incompleteness, to the model.

\begin{figure}
\plotone{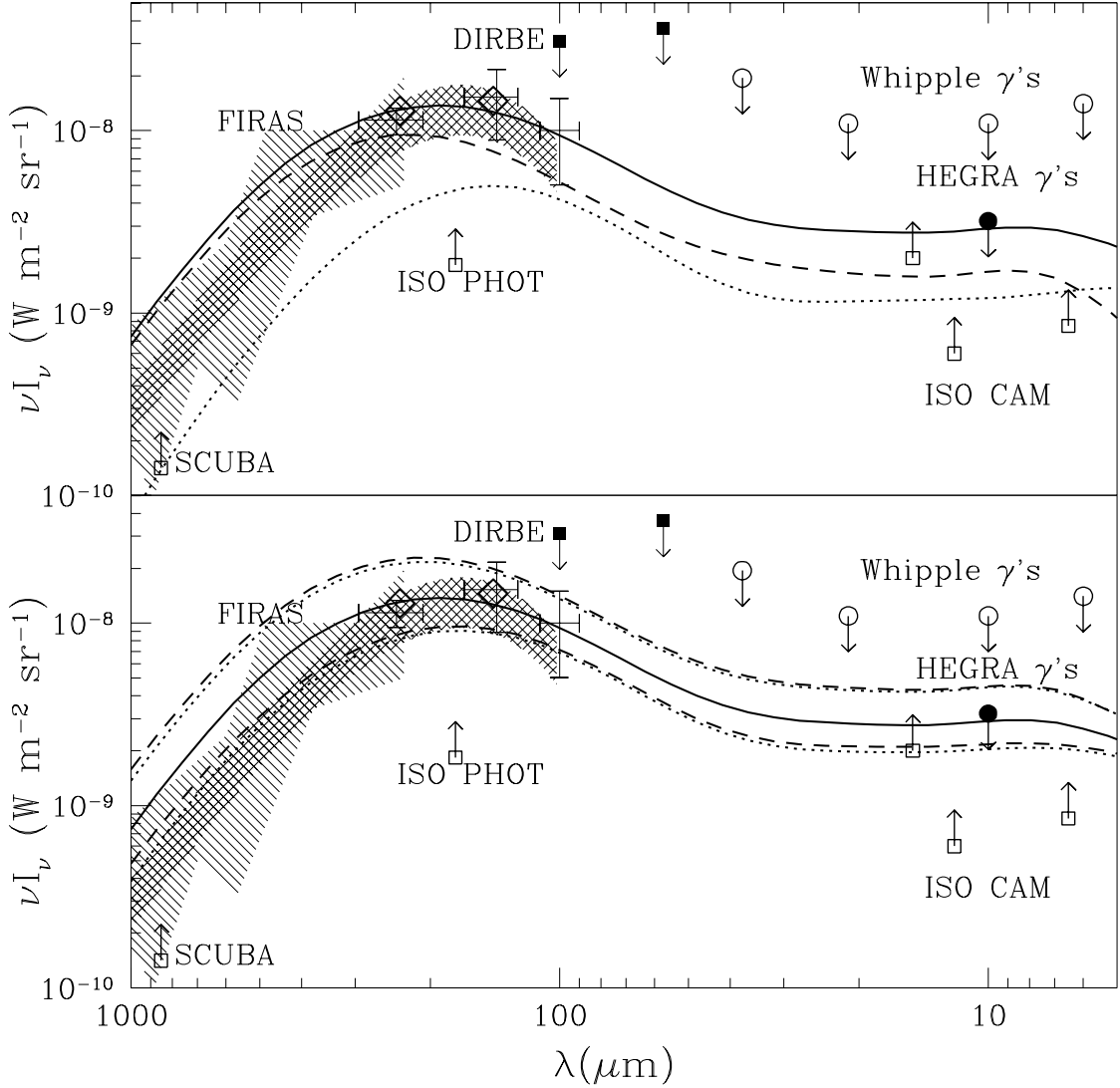}
\caption{ Extra-galactic background. The model is accurate for
wavelengths longer than $9\:{\rm \mu m}$. At shorter wavelengths the
model result should be regarded as a lower limit.  {\it Top Panel:}
Fiducial model results: {\it dotted line} - disks, {\it dashed line} -
starbursts, {\it solid line} - total. {\it Bottom Panel:} Effect of
varying model parameters on the {\it total} background: {\it solid
line} - fiducial model, {\it upper dotted line} - $L_{cut}=11.2\times
10^{10}\:h^{-2}\:L_{\odot}$, {\it lower dotted line} -
$L_{cut}=2.8\times 10^{10}\:h^{-2}\:L_{\odot}$, {\it upper dashed
line} - $f_{0}^{gas}=0.05$, {\it lower dashed line} -
$f_{0}^{gas}=0.2$.  Data : {\it diagonally-shaded region} - FIRAS -
$1\:\sigma$ interval from Puget et al (1996), {\it hatched-shaded
region} - FIRAS - Fixsen et al (1998), {\it Crosses} - DIRBE - Lagache
et al (1999) and Dwek et al (1998) (at $100\: {\rm \mu m}$), {\it Open
diamonds} - Fixsen et al (1998), {\it Solid squares} - Schlegel et al
(1998), {\it Open circles} - Biller et al (1998) from observations of
TeV $\gamma$-rays (Whipple collaboration), {\it Solid circle} - Stanev
\& Franceschini (1998) from HEGRA $\gamma$-ray observations, {\it Open
squares} - number count limits from SCUBA and ISO.
\label{fig:ebl}}
\end{figure}

The predictions of the extra-galactic background in the IR to sub-mm
are shown in Figure \ref{fig:ebl}. Recent independent estimates of the
flux from the FIRAS residuals between 150 - 5000 ${\rm \mu m}$ (Puget
et al 1996; Fixsen et al 1998) agree on the spectrum and amplitude of
the background, although the sizes of the systematic errors are not
well determined. Strong upper limit constraints are also being
reported at shorter wavelengths from observations of TeV $\gamma$-rays
(Stanev \& Franceschini 1998; Biller et at 1998), while lower limits
can be placed from the summed flux predicted by number count
surveys. Our fiducial model is consistent with the existing data,
although at longer wavelengths it predicts a flux about twice as high
as the mean of the FIRAS amplitudes. This discrepancy is within the
bounds of model and observational uncertainties, as illustrated in the
lower panel of Figure \ref{fig:ebl}.

\begin{figure}
\plotone{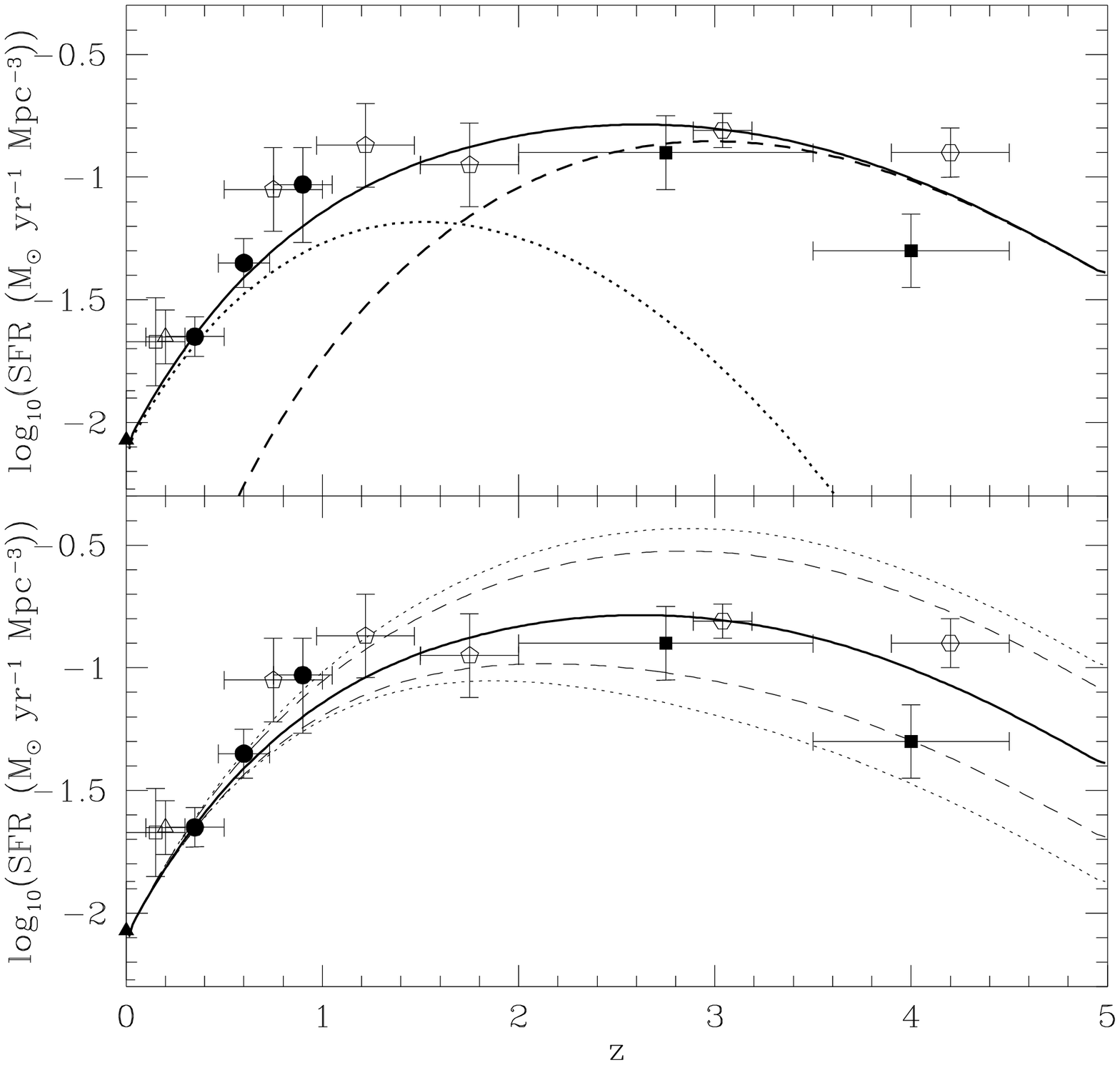}
\caption{ Global Star Formation Rate History.
{\it Top Panel:} Fiducial model results: {\it dotted line} -
contribution from disk galaxies. {\it dashed line} - contribution form
starburst galaxies. {\it solid line} - total.
{\it Bottom Panel:} Effect of varying $L_{cut}$ and $f_{0}^{gas}$ on
the {\it total} global SFR: {\it solid line} - fiducial model, {\it
upper dotted line} - $L_{cut}=11.2\times 10^{10}\:h^{-2}\:L_{\odot}$,
{\it lower dotted line} - $L_{cut}=2.8\times
10^{10}\:h^{-2}\:L_{\odot}$, {\it upper dashed line} -
$f_{0}^{gas}=0.05$, {\it lower dashed line} - $f_{0}^{gas}=0.2$.
Data: {\it solid triangle} from Gallego et al (1995), {\it open square} from
Treyer at al (1998), {\it open triangle} from Tresse \& Maddox (1998),
{\it solid circles} from Lilly et al (1996), {\it open pentagons} from
Connolly et al (1997), {\it solid squares} from Madau et al (1996) and
{\it open hexagons} Steidel et al (1998).  The data beyond $z\sim0.3$
are extinction corrected with the Calzetti (1997) reddening law, following Steidel et al (1998).
\label{fig:globalSFR}}
\end{figure}

Finally we derive the global SFR history (Figure \ref{fig:globalSFR})
by assuming a conversion factor of $2\times10^{-10}\:{\rm
M_{\odot}\:yr^{-1}\:L^{FIR}_{\odot}}$. This is in agreement with
recent calibrations ($1 - 3\times10^{-10}\:{\rm
M_{\odot}\:yr^{-1}\:L^{FIR}_{\odot}}$) from starburst synthesis models
(Leitherer \& Heckman 1995; Lehnert \& Heckman 1996; Meurer et al
1997). This conversion factor is applied to the far-IR luminosity due
to dust heating by young stars in disks, accounting for the evolving
optical depth, as well as to the optically thick starburst population.
The results for the disk contribution are thus relatively sensitive to
our choice of $\tau_{0}^{uv}\sim0.7$. However, with this fiducial
value, the model predictions agree well with the low redshift
observations, given the uncertainties in the SFR to far-IR luminosity
conversion factor. At high redshifts the model is consistent with
recent results which suggest a relatively flat SFR history out to
$z\sim4$ (Steidel et al 1998).

If we assume that our starbursts lead to the formation of spheroidal
stellar systems, such as ellipticals and the bulges and halos of disk
galaxies, then we can predict the mass densities of the disk and
spheroidal components that exist in the present-day
Universe. Neglecting mass returned to the inter-stellar and
inter-galactic media, we find $\Omega_{disk}\approx5.1\times10^{-3}$
and $\Omega_{sph}\approx4.4\times10^{-3}$. These estimates are
sensitive to $L_{cut}$ and are uncertain by factors of a
few. Observationally, the spheroidal component is dominated by the
bulges and halos of disk galaxies, due to the paucity of ellipticals
in the field (Binggeli, Sandage \& Tammann 1988). Our results suggest
the mass in spheroids is comparable to the mass in the disks,
consistent with the analysis of Schechter \& Dressler (1987).

Our model indicates that there is a broad peak of spheroid formation
at $z\sim3$ (Figure \ref{fig:globalSFR}). Recently, evolved
ellipticals have been seen in deep NICMOS images (Benitez et al 1998),
requiring a very high ($z\gtrsim5$) redshift of formation.
Extrapolating the global SFR predictions of our model to these high
redshifts, whilst uncertain, indicates that the SFR's are high enough
to account for some very early elliptical formation. Note, however,
within the framework of our model, the contribution to the
extra-galactic background and the number counts from these early times
($t<1$ Gyr) is negligible. In other words, almost all the star
formation and associated energy release necessary to account for the
observed extra-galactic background occurs at redshifts $\lesssim 5$
and is accessible to current observations.

\section{Conclusions}

We have presented a simple model for the evolution of disk galaxies
and starbursts, tied as closely as possible to observations. Despite
its simplicity, the model takes account of dust, gas and spectral
evolution in a self-consistent manner, and is able to predict source
counts and the extra-galactic background in the IR to sub-mm
consistent with observations, without recourse to fine tuning of
parameters. A disk-only model, with $L_{cut}=L_{max}$, gives only
marginally greater counts and fluxes than the disk component of the
fiducial model, thus failing to account for observations. This
demonstrates the significant role of the starburst population. The
fiducial model we present is based on the best estimates and
observations of our various model parameters (Table
\ref{tab:param}). We have not attempted to vary these to obtain the
best fit to the, often uncertain, high redshift IR and sub-mm data.

The predicted global SFR history agrees with recent observations,
corrected for dust extinction, indicating that SFR's remain high from
$z\sim1$ back to $z\gtrsim4$. At high redshifts the SFR is dominated
by starbursts and is driven by galaxy-galaxy interactions. Over the
history of the Universe, the total star formation occurring in
starbursts is comparable to that in disks. This suggests the
baryonic mass in bulges, formed from starbursts, is similar to that in
disks. The bulges form at high redshift, and are then thought to act
as the seeds for disk formation as gas infalls.

The energy release associated with the entire star formation history
to $z\sim5$ is enough to account for all of the observed
extra-galactic background. This implies that the vast majority of star
formation in the Universe occurs over this period.

The majority ($\sim 2/3$) of the sources recently detected with SCUBA
are identified as starbursts. Typical redshifts are $1\lesssim z
\lesssim 5$.  The direct effect of these sources, along with other
foregrounds, on the angular power spectrum of the microwave background
has been examined by Gawiser et al (1999). The effect of
cluster-induced lensing of the sources on the background has been
investigated by Scannapieco, Silk \& Tan (1999) and found to be small.

Future possible improvements include a more sophisticated method of
distinguishing disks and starbursts in the infra-red luminosity
function, a more detailed treatment of the opacity of starbursts and
the inclusion of active galactic nuclei. Extending the model to the
ultra-violet, optical and radio, will allow additional observations to
help constrain the galaxy evolution. Consideration of alternative
cosmologies is left to a future date. The model will be considerably
refined once results are available from future observations with
FIRST, SIRTF and the NGST. Model results are available electronically
at http://astro.berkeley.edu/~jt/irmodel.html.

\acknowledgements We are grateful to Matt Malkan for kindly providing
his data for the IR galaxy spectra in electronic form, and to Herve
Aussel for his ISO CAM 15 ${\rm \mu m}$ number count data. We thank
R. Bouwens, J. Puget, E. Scannapieco, A. Cumming and E. Gawiser for
helpful discussions.

\clearpage

\begin{deluxetable}{cccc} 
\small
\tablecaption{Model Parameters\label{tab:param}}
\tablewidth{0pt}
\tablehead{
\colhead{Parameter} & \colhead{Fiducial Value} & \colhead{Description \& Reference} & \colhead{Model Sensitivity}\nl
}
\startdata
$L_{cut}$ & $5.6\times10^{10}\:h^{-2}\:L_{\odot}$ & Present-day disk-starburst division & high\\
 & & - Sanders \& Mirabel (1996) & \\
\\
$L_{min}$ & $1.0\times 10^{5}\:h^{-2}\:L_{\odot}$ & Present-day min disk luminosity & low\\
 & & - Saunders et al (1990) & \\
\\
$L_{max}$ & $1.0\times 10^{15}\:h^{-2}\:L_{\odot}$ & Present-day max starburst luminosity & low\\
 & & - Kim \& Saunders (1998) & \\
\\
$f_{0}^{gas}$ & 0.1 & Gas fraction of baryons bound to disks & high\\
 & & - Young et al (1995) & \\
\\
$f_{0}^{young}$ & 0.65 &Fraction of IR $L_{*}^{disk}$ heated by young& low\\
 & &stars - Mayya \& Rengarajan (1997)& \\
\\
$\tau_{0}^{uv}$ & 0.7 & Disk optical depth to young stellar & low\\
 & & energy release - Wolfire et al (1999) & \\
\\
$\tau_{0}^{opt}$ & 0.1 & Disk optical depth to old stellar & low\\
 & & energy release - Wolfire et al (1999) & \\
\\
$z_{f}$ & 5.5 & Galaxy formation redshift from age & low\\
 & & of Milky Way - Knox et al (1999) & \\
\\
$\beta$ & 1.5 & Dust emissivity, $\nu^{\beta}$ & low\\
 & & - Franceschini et al (1998) & \\
\\
$n$ & 1.5 & Schmidt law index, $\phi\propto(\rho^{gas})^{n}$ & low\\
 & & - Kennicutt (1998) & \\
\\
\enddata
\end{deluxetable}

\begin{references}
\noindent
Altieri, B., et al 1998, accepted by \aap, astro-ph/9810480\\
Aussel, H., Cesarsky, C.J., Elbaz, D., \& Starck, J.L. 1998, submitted to \aap, astro-ph/9810044\\
Balland, C., Silk, J., \& Schaeffer, R. 1998, \apj, 497, 541\\
Benitez, N., Broadhurst, T., Bouwens, R., Silk, J., \& Rosati, P. 1998, submitted to \apjl, astro-ph/9812205\\
Biller, S.D., et al. 1998, \prl, 80, 2992\\
Binggeli, B., Sandage, A., \& Tammann, G.A. 1988, \araa, 26, 509\\
Blain, A.W. \& Longair, M.S. 1993, \mnras, 264, 509\\
Blain, A.W., Kneib, J.-P., Ivison, R.J., Smail, I., 1999, \apjl, submitted\\
Burigana, C. \& Popa, L. 1998, \aap, 334, 420\\
Calzetti, D. 1997, \aj, 113, 162\\
Connolly, A.J., Szalay, A.S., Dickinson, M., Subbarao, M. U., \& Brunner, R.J. 1997, \apj, 486, L11\\
Devereux, N.A., Price, R., Wells, L.A., \& Duric, N. 1994, \aj, 108, 1667\\
Dwek, E., et al 1998, submitted to \apj, astro-ph/9806129\\
Edvardsson, B., Anderssen, J., Gustafsson, B., Lambert, D., Nissen, P., \& Tomkin, J. 1993, \aap, 275, 101\\
Fisher, K.B., Strauss, M.A., Davis, M., Yahil, A., \& Huchra, J.P. 1992, \apj, 389, 188\\
Fixsen, D.J., Dwek, E., Mather, J.C., Bennett, C.L., Shafer, R.A. 1998, submitted to \apj, astro-ph/9803021\\
Franceschini, A., Andreani, P., \& Danese, L. 1998, \mnras, 296, 709\\
Franceschini, A., Mazzei, P., De Zotti, G., Danese, L. 1994, \apj, 427, 140\\ 
Gallego, J., Zamorano, J., Aragon-Salamanca, A., \& Rego, M. 1995, \apjl, 361, L1\\
Gawiser, E., et al 1999, in preparation. See also http://astro.berkeley.edu/wombat\\
Gispert, R., Puget, J.L., \& Lagache, G. 1999, in preparation\\ 
Guiderdoni, B., Hivon, E., Bouchet, F.R., \& Maffei, B. 1998, \mnras, 295, 877\\
Hauser, M.G., et al 1998, submitted to \apj, astro-ph/9806167\\
Jimenez, R. \& Kashlinsky, A. 1998, accepted to \apj, astro-ph/9802337\\
Kennicutt, R.C. 1998, \apj, 498, 541\\
Kauffmann, G. and Charlot, S. 1998, MNRAS, 283, L117\\
Kim, D.-C., \& Saunders, D.B. 1998, submitted to \apjs, astro-ph/9806148\\
Knox, R.A., Hawkins, M.R.S., \& Hambly, N.C. 1999, accepted to \mnras, astro-ph/9903345\\
Lagache, G., Abergel, A., Boulanger, F., Desert, F.X., \& Puget, J.L. 1999, accepted to \aap, astro-ph/9901059\\
Lehnert, M., \& Heckman, T. 1996, \apj, 472, 546\\
Leitherer, C., \& Heckman, T.M. 1995, \apjs, 96, 9\\
Lilly, S.J., Le Fevre, O., Hammer, F., \& Crampton, D. 1996, \apj, 460, L1\\
Lonsdale, C.J., Hacking, P.B., Conrow, T.P., \& Rowan-Robinson, M. 1990, \apj, 358, 60\\
Madau, P., et al 1996, \mnras, 283, 1388\\
Malkan, M.A., \& Stecker, F.W. 1998, \apj, 496, 13\\
Meurer, G.R., Heckman, T.M., Lehnert, M.D., Leitherer, C., \& Lowenthal, J. 1997, \aj, 114, 54\\
Oswalt, T.D., Smith, J.A., Wood, M.A., \& Hintzen, P. 1996, \nat, 382, 692\\
Pei, Y.C., Fall, S.M., \& Hauser, M.G. 1998, submitted to \apj, astro-ph/9812182\\
Prantzos, N., \& Silk, J. 1998, \apj, 507, 229\\
Puget, J.L., et al 1996, \aap, 308, L5\\
Puget, J.L., et al 1998, accepted to \aap, astro-ph/9812039\\
Rocha-Pinto, H., \& Maciel, W. 1996, \mnras, 279, 447\\
Roche, P.F., \& Chandler, C.J., 1993, \mnras, 265, 486\\
Rush, B., Malkan, M., Spinoglio, L. 1993, \apjs, 89, 1\\ 
Sanders, D.B., \& Mirabel, I.F. 1996, \araa, 34, 749\\
Saunders, W., et al 1990, \mnras, 242, 318\\
Sackett, P.D. 1997, \apj, 483, 103\\
Scannapieco, E., Silk, J., \& Tan, J.C. 1999, submitted to \apj\\
Schechter, P.L., \& Dressler, A. 1987, \aj, 94, 563\\ 
Schlegel, D.J., Finkbeiner, D.P., \& Davis, M. 1998, \apj, 500, 525\\
Smail, I., Ivison, R., Blain, I. \& Kneib, J.P. 1998, Invited review at the Maryland Astrophysics Conference ``After the dark ages: when galaxies were young (the Universe at $2<z<5$)'', astro-ph/9810281\\
Stanev, T., \& Franceschini, A. 1998, \apj, 494, 159\\
Steidel, C.C., Adelberger, K.L., Giavalisco, M., Dickinson, M., \& Pettini, M., 1998, \apj, submitted, astro-ph/9811399\\
Trentham, N., Blain, A.W., \& Goldader, J. 1999, \mnras, submitted, astro-ph/9812081\\
Tresse, L. \& Maddox, S. 1998, \apj, 495, 691\\
Treyer, M.A., Ellis, R.S., Milliard, B., Donas, J., \& Bridges, T.J. 1998, \mnras, accepted, astro-ph/9806056\\
Walterbos, R.A.M., \& Greenawalt, B., 1996, \apj, 460, 696\\
Williams, R.E., et al 1996, \aj, 112, 1335\\
Wolfire, M.G., et al 1998, in preparation\\
Wyse, R., \& Gilmore, G. 1995, \aj, 110,2771\\
Xu, C., \& Helou, G., 1996, \apj, 456, 152\\
Young, J.S., \& Scoville, N.Z., 1991, \araa, 29, 581\\
Young, J.S., et al 1995, \apjs, 98, 219\\
Zepf, S. 1997, Nature, 390, 377\\
\end{references}
\end{document}